\newcommand{\ket}[1]{\lvert #1\rangle}

\newcommand{\Tr}{\operatorname{Tr}}
\newcommand{\Jx}{J_{ex}}

\def\Q{{\@QC Q}}
\def\C{{\@QC C}}
\def\@QC#1{\mathpalette{\setbox0=\hbox\bgroup$\rm}%
  {\egroup C$\egroup\rm\rlap{\kern0.4\wd0\vrule
  width 0.05\wd0 height 0.97\ht0 depth -0.01\ht0}%
  #1\bgroup}}

\documentclass[aps,twocolumn,raggedbottom,prb,showpacs,nobalancelastpage,amssymb,groupedaddress]{revtex4}
\usepackage{color}
\usepackage{graphicx}
\usepackage{amsmath}
\usepackage{amssymb}
\include{nem}

\begin{document}
\title{Swapping and entangling hyperfine coupled nuclear spin baths}
\author{B.\, Erbe and J.\, Schliemann}
\affiliation{Institut f\"{u}r Theoretische Physik, Universit\"at
Regensburg, 93053 Regensburg, Germany}
\date{\today}

\begin{abstract}
We numerically study the hyperfine induced nuclear spin dynamics in a system 
of two 
coupled quantum dots in zero magnetic field. Each of the electron spins is 
considered to interact with an individual bath of nuclear spins via 
homogeneous coupling constants (all coupling coefficients being equal). 
In order to lower the dimension of the problem, the two baths are approximated 
by two single long spins. 
We demonstrate that the hyperfine interaction enables to utilize the nuclear 
baths for quantum information purposes. 
In particular, we show that it is possible to swap the nuclear ensembles on 
time scales of seconds and 
indicate that it might even be possible to fully entangle them. 
As a key result, it turns out that
the larger the baths are, the more useful they become as a resource of 
quantum information. Interestingly, the nuclear spin
dynamics strongly benefits from combining two  quantum dots of 
\textit{different} geometry to a double dot set up.
\end{abstract}
 \maketitle

\textit{Introduction.}--Electron spins confined in semiconductor 
quantum dots with an s-type 
conduction
band, like for example GaAs quantum dots, experience decoherence 
through the 
spin-orbit
interaction,  and by the hyperfine interaction with surrounding 
nuclear spins.
With respect to possible future solid state quantum computation systems
utilizing the electron spin as the qubit \cite{LossDi98,Hanson07}, these 
interactions act as a source of
decoherence. Due to the spatial confinement of the electron spin in a quantum 
dot, the relaxation time $T_1$ induced 
by the spin-orbit interaction is enhanced for low temperatures 
\cite{KhaNaz00, KhaNaz01}. As the dephasing time $T_2$ due to the spin orbit 
interaction
turns out to be as long as the $T_1$ time under realistic conditions 
\cite{GolKhaLoss04}, the major source of decoherence in semiconductor 
quantum dots results from the hyperfine interaction 
\cite{KhaLossGla02,KhaLossGla03,Petta05,Koppens06,Braun05}. 
For related reviews the reader is referred to Refs. 
\cite{SKhaLoss03,Zhang07,Klauser07,Coish09,Taylor07}. 
Similar situations arise in carbon nanotube quantum dots \cite{Church09}, 
phosphorus donors in silicon \cite{Abe04} and nitrogen vacancies in diamond 
\cite{Jel04, Child06, Hanson08}.

Apart from this detrimental effect of the hyperfine interaction, 
it provides a way to efficiently access the nuclear spins by e.g. external 
degrees of freedom. This for example enables to built up an interface 
between light and nuclear spins \cite{SchCiGi08,SchCiGi09}, to polarize 
nuclear spin baths \cite{Taylor03,ChriCiGi09, ChriCiGi07}, to set up 
long-lived quantum \cite{Taylor032,Morton08} and classical \cite{Austing09} 
memory devices or to generate entanglement \cite{ChriCiGi08}. 

In both of the aforementioned contexts it is of key importance to understand 
the hyperfine
induced spin dynamics. Here one has to distinguish between the case of a
strong and the case of a weak 
magnetic field
applied to the electron spins. In the first limit,  the ``flip-flop''
terms between the electron and the nuclear spins occurring in the Hamiltonian
are strongly suppressed. This allows to treat them perturbatively or to even
completely neglect them, which strongly simplifies the calculations 
\cite{KhaLossGla02, KhaLossGla03, Coish04, Coish05, Coish06, Coish08}.
In the absence of such an external magnetic field, however, many 
approximative techniques break down, and one has to resort to exact methods. 
As explained in Ref. \cite{ErbSchl09}, in order to gain exact results, 
strong restrictions 
on the initial state \cite{KhaLossGla02,KhaLossGla03}, the size of the 
system \cite{SKhaLoss02, SKhaLoss03} or the hyperfine coupling constants 
\cite{BorSt07, ErbS09, ErbSchl09, ErbSchl10} have to be made. 

In the present paper we combine the second and the third approach and 
focus on the \textit{advantages}
of the hyperfine interaction. To this end we consider a model of 
two exchange coupled electron spins each of which is interacting with an 
\textit{individual} bath of nuclear spins. 
This corresponds to the situation of spatially well-separated quantum dots.
Assuming the baths to be strongly
polarized in opposite directions initially, we investigate, by means of 
exact numerical 
diagonalization, to what extent it is possible to swap and entangle them. 
Usually exact numerical diagonalizations are restricted to rather small
 system 
sizes. In order to go beyond these limits, we reduce the dimension of the 
problem by approximating the two baths by two single long spins. The 
spectral properties of the model described above have recently
been studied in Ref. \cite{ErbSchl10}, where it has been shown that the 
spectrum of the Hamiltonian 
exhibits systematically degenerate multiplets. Motivated by these findings, 
below we will distinguish between an inversion symmetric system, showing 
the mentioned degeneracies, and a system with broken inversion symmetry 
where such degeneracies are absent.

The work presented in this paper 
complements the results of Ref. \cite{ErbSchl09}, where we analytically 
studied the homogeneous coupling case for two electron spins coupled to a 
common bath of nuclear spins. Choosing the hyperfine coupling constants 
to be equal to each other of course has to be regarded as a rough 
approximation. However, it has been shown that already such simple models can 
yield concrete predictions and realistic results 
\cite{ErbSchl09,SchCiGi08,SchCiGi09,ChriCiGi09,ChriCiGi08}.
 
\textit{Model and methods.}--The Hamiltonian of two exchange coupled electron 
spins, each of which is interacting with an
individual bath of nuclear spins via the hyperfine interaction reads
\begin{equation}
\label{1}
 H= \vec{S}_1 \cdot \sum_{i=1}^{N_1} A_i^1 \vec{I}_{i1} 
+ \vec{S}_2 \cdot \sum_{i=1}^{N_2} A_i^2 \vec{I}_{i2} + \Jx \vec{S}_1 
\cdot \vec{S}_2 ,
\end{equation}
where $\vec{S}_j$ are the electron and $\vec{I}_{ij}$ are the nuclear 
spins the $j$-th electron spin interacts with. For
simplicity we will consider $N_1=N_2=:N$ in what follows. The 
parameter $\Jx$ denotes an exchange coupling between the two electron 
spins, which can experimentally be adjusted in a range of 
$[-10^{-3}, 10^{-3}]$eV, and 
$A_i^1 $, $A_i^2 $ are the hyperfine coupling constants. In a realistic 
quantum dot these are proportional to the square modulus of the
electronic wave function at the sites of the surrounding nuclear spins. 
As a typical example, in GaAs quantum dots the overall coupling strength 
of the $j$-th electron spin $A^j:=\sum_{i=1}^N A_i^j$ is of 
the order of $\left[ 10^{-4},10^{-5}\right] $eV.

Due to the spatial variation of the electronic wave function, the 
hyperfine couplings are clearly inhomogeneous.
However, in the following we consider them to be equal to each other,
meaning that $A^j_i=A^j/N$. Then the Hamiltonian (\ref{1}) 
conserves apart from the total spin 
$\vec{J}=\vec{S}_1+\vec{S}_2+\vec{I}_1 + \vec{I}_2$, 
where
$\vec{I}_j=\sum_{i=1}^{N_j}\vec{I}_{ij}$, also the squares of the total 
bath spins $\vec{I}^2_j$
\begin{equation}
\label{symm}
\left[H,\vec{J}\right]=\left[H, \vec{I}_j^2\right]=0.
\end{equation}
The first symmetry will be helpful for the exact numerical diagonalizations of the 
Hamiltonian
matrix \cite{SKhaLoss02,SKhaLoss03}, through which we will obtain the 
dynamics in the following. We
compute the time dependent density matrix by decomposing the initial 
state into energy eigenstates and applying
the time evolution operator. Tracing out the electron degrees of freedom 
then yields the reduced density matrix of the nuclear baths $\rho_n(t)$ 
from which we can calculate the time evolution of all observables. For 
details the reader is referred to Ref. \cite{ErbSchl09}.

In the following we will approximate each of the two baths $\vec{I}_j$ by 
a single long
spin. Let us briefly discuss to which physical situation this corresponds. 
A general state of a bath is a superposition of states from different multiplets
\begin{equation}
\label{Bstate}
 \ket{\beta_j}=\sum_{I_j,m_j} \beta_j^{I_j,m_j} \ket{I_j,m_j},
\end{equation}
where the quantum numbers associated with a certain Clebsch-Gordan decomposition 
of the bath have been omitted. The number
of multiplets contributing to the sum in (\ref{Bstate}) decreases
with increasing bath polarization. For very high polarizations we can therefore approximate 
the state (\ref{Bstate}) by $\ket{\beta_j}=\ket{I,m_j}$ with $m_j \approx \pm I$.
Due to the commutation relations (\ref{symm}) all dynamics is then captured by the following simple Hamiltonian, 
to which we refer to as the long spin approximation Hamiltonian:
\begin{equation}
\label{Ham}
 H_{\text{LSA}}=\frac{A^1}{2I} \vec{S}_1 \cdot \vec{I}_1 
+ \frac{A^2}{2I} \vec{S}_2 \cdot \vec{I}_2 + J_{ex} \vec{S}_1 \cdot \vec{S}_2
\end{equation}
The form of the couplings $A^j/2I$ results from the observation that 
the $N$ bath spins can couple to
$I=N/2, N/2-1, N/2-2, \ldots$. As we assume highly polarized baths, 
we consider the maximal value $I=N/2$. 
Solving for $N$ then yields the coupling constants in (\ref{Ham}). 
For later convenience we define $A:=A^1+A^2$.
\begin{figure}
\begin{flushright}
\resizebox{\linewidth}{!}{
\includegraphics{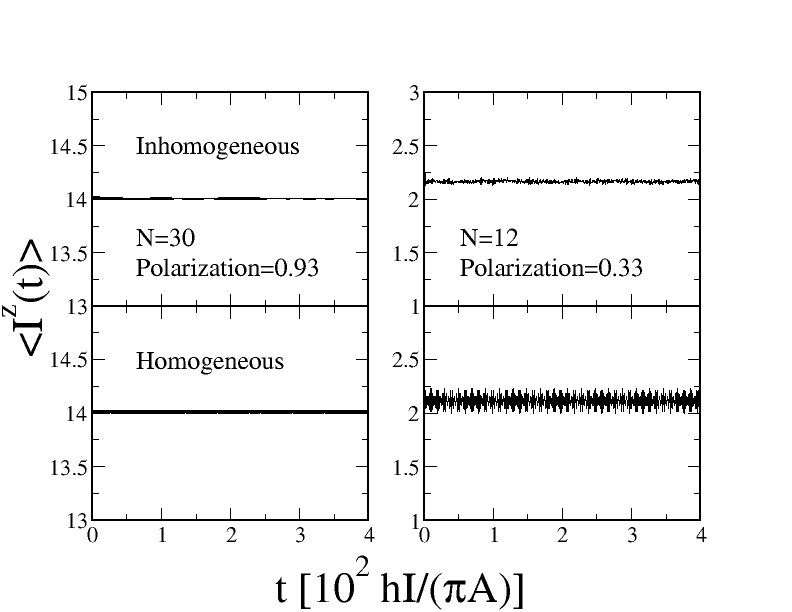}}
\end{flushright}
\caption{\label{nucl} Nuclear spin dynamics of the Gaudin model for inhomogeneous (upper panels)
and homogeneous (lower panels) hyperfine couplings. The bath state is
initially randomly correlated \cite{SKhaLoss02,SKhaLoss03} and the electron spin is
pointing upwards. The left column shows the case of $N=30$ and a bath 
polarization of $0.93$, while in the right column we have $N=12$ and a
polarization of $0.33$. Even in the latter case the bath dynamics for 
inhomogeneous and homogeneous couplings are still quite similar to each other.}
\end{figure}

High nuclear polarizations of up to $80 \%$ have been experimentally 
demonstrated in Refs.~\cite{Atac,Ono,Baugh,Tarucha}. In particular,  
Ref.~\cite{Tarucha} also discussed the possibility of polarizing two 
nuclear ensembles in different directions.
However, a question concerning the LSA arises from assuming the couplings 
to be homogeneous:
As demonstrated in Ref.~\cite{ErbSchl09}, this approximation is a good
one for short
time scales, whereas for longer times artifacts occur. As the nuclear 
dynamics are slow, it has to 
be questioned to what extent homogeneous couplings are adequate in order to 
evaluate nuclear spin
dynamics. Therefore we numerically investigated the time evolution
of the nuclear spins in the Gaudin model. The Gaudin model is the
central spin model with a single central spins, as corresponding to one of 
the two first terms in the Hamiltonian (\ref{1}). As a result from our numerics, we find that the influence of 
inhomogeneities is suppressed with increasing polarization. This is
illustrated Fig.~\ref{nucl} where we compare cases of high and 
low polarization. Even in the latter case the dynamics for both types of 
couplings are very close to each other. In the LSA we are considering
two coupled Gaudin models. Consequently, the data presented in Fig. \ref{nucl}
does not give a strict proof for the adequacy of neglecting inhomogeneities
in the couplings. However, it clearly provides qualitative evidence into this direction.

\textit{Swapping nuclear spin polarizations}.--We will now evaluate the nuclear 
spin dynamics within the LSA and explore the possibility of swapping
oppositely polarized spin baths, $\langle I_1^z \rangle =- \langle I_2^z \rangle$.  Our initial state 
$\ket{\alpha}$ will be a simple product state between the electron and 
the nuclear state $\ket{\alpha}=\ket{\alpha_e} \ket{\alpha_n}$.
Since both baths are spatially well-separated, the initial nuclear state
is again a product state of the two long spins. Note that entanglement
within the baths can not be considered within the LSA. However, due
to their high polarizations, correlations within the baths
are of minor importance. In what follows, we will 
always work in subspaces of fixed $J^z=:M$ where only the $z$-component
has a non-zero expectation value $\langle I_j^z(t) \rangle$. Note that
(again) due to the high bath polarizations, it is realistic to assume that the
initial state has components exclusively in subspaces of fixed $M$.
Moreover, we will assume the $z$-component of the total electron spin to
be initially zero, i.e. the spins are antiparallel. Similar results as to be 
presented below are obtained for more general initial states of the
electron spin system.

\begin{figure}
\begin{flushright}
\resizebox{\linewidth}{!}{
\includegraphics{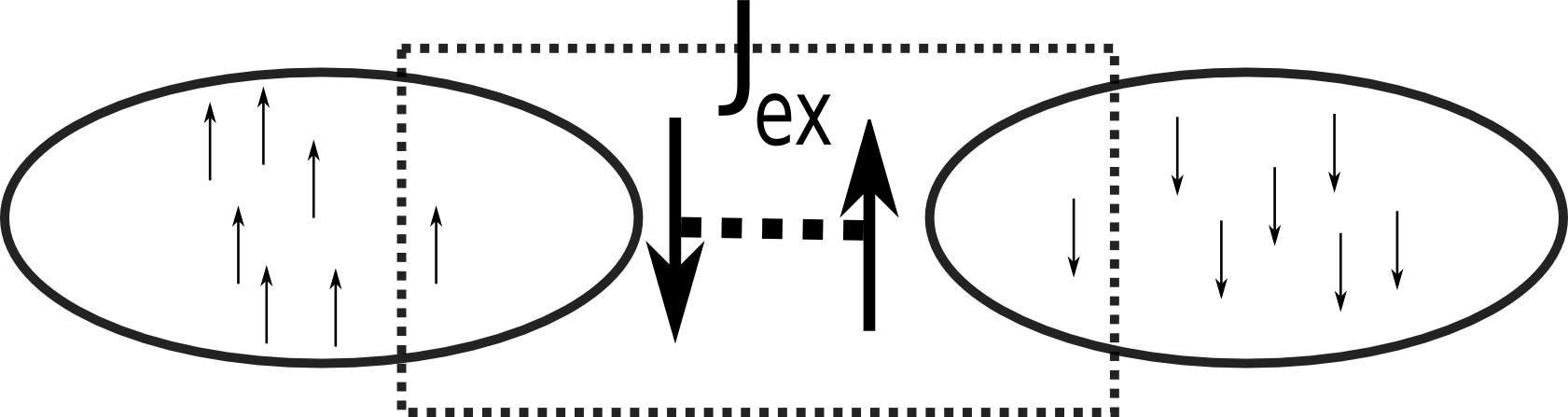}}
\end{flushright}
\caption{\label{bath} The nuclear baths are considered to consist of spins 
with length $(1/2)$. The frame marks the LSA system with the smallest 
possible bath spin length $I=1/2$. Here the dynamics of all four spins are highly 
coherent provided the values of all couplings are close to each other,
motivating the condition (\ref{coupling}) (see text).}
\end{figure}

\begin{figure}
\begin{flushright}
\resizebox{\linewidth}{!}{
\includegraphics{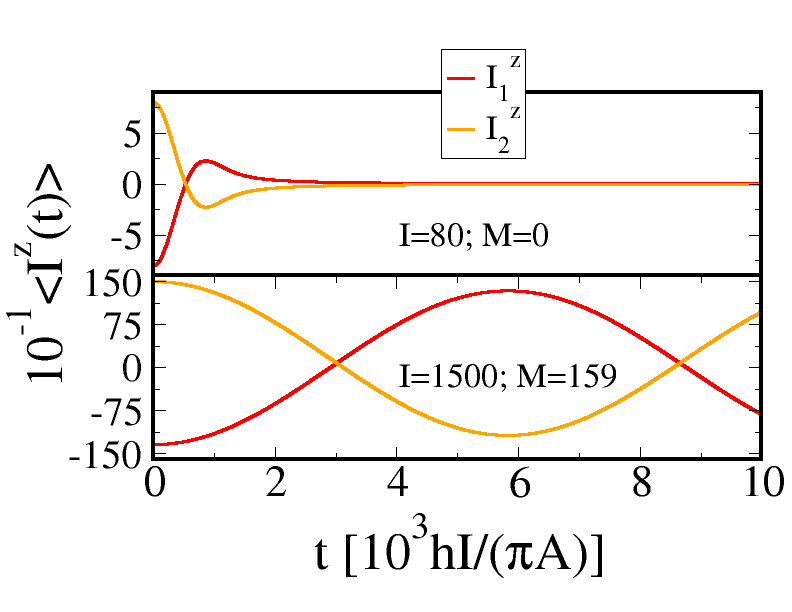}}
\end{flushright}
\caption{\label{No} Nuclear spin dynamics for $\Jx/(A/2I)=3.5$, $\Delta=0$ 
and $\ket{\alpha}=(1/\sqrt{13})\left(2 \ket{\Uparrow \Downarrow}
+3\ket{\Downarrow \Uparrow} \right)\ket{M-I,I}$. The upper panel shows data 
data $I=80$ and $M=0$ where the expectation values $\langle I_j^z(t) \rangle$
decay rapidly to zero. In the bottom panel we consider
$I=1500$ and $M=159$ where an almost complete swap of the nuclear spins
is observed.}
\end{figure}

More importantly, we will concentrate on exchange couplings
being of the same order of magnitude
as the hyperfine coupling strength,
\begin{equation}
\label{coupling}
\Jx /(A/2I) \approx 1, 
\end{equation}
meaning that the two electron spins are coupled as strongly to each other as 
they are coupled to the bath spins. This is motivated by the following 
observation. Let us consider (\ref{Ham}) for the smallest possible value, 
$I=1/2$. As shown by elementary numerics, the dynamics of all four spins
are, under the condition (\ref{coupling}), highly coherent, and the
nuclear spin polarizations can nicely be swapped, i.e. at the end of
the process the expectation values $\langle I_j^z(t) \rangle$ are, to a very
good degree of accuracy, exchanged as compared to the initial state.
Let us now consider the two baths in the original model (\ref{1}) to 
consist of spins 
with length $(1/2)$, as already assumed for the derivation of the 
couplings in (\ref{Ham}).
As depicted in Fig. \ref{bath}, the complete system can now be regarded 
as set of $I=1/2$ models. Thus, from a heuristic point of view, 
the biggest chance to swap the full baths exists if all the subsystems 
are swapped. Hence, the exchange coupling has to be of the order of the 
coupling between the electron and the bath spins for \textit{any} subsystem. 
For homogeneous couplings within the baths, this means that $\Jx \approx A/N$, 
which translates into (\ref{coupling}) for the LSA, as explained above.

However, at first sight, it 
does not seem to be possible to swap the initially 
antiparallel nuclear spins $I>1/2$, even if the condition (\ref{coupling})
is fulfilled. This is demonstrated in the 
upper panel of Fig. \ref{No}, where we consider $I=80$, $\Jx/(A/2I)=3.5$ and 
a zero ``detuning'' $\Delta:=A_2-A_1=0$. This corresponds
to a situation in which the two quantum dots have the same geometries. 
We choose the comparatively generic 
electron spin state 
$\ket{\alpha_e}=(1/\sqrt{13})\left(2 \ket{\Uparrow \Downarrow}
+3\ket{\Downarrow \Uparrow} \right)$ 
(similar results occur for other choices)
and plot the dynamics for 
antiparallel nuclear spin configurations with the maximal possible $z$ 
components $\ket{\alpha_n}=\ket{-I,I}$. As seen from the figure, the expectation
values of the bath spins decrease quite rapidly to zero and are far away from 
being properly swapped.

\begin{figure}
\begin{flushright}
\resizebox{\linewidth}{!}{
\includegraphics{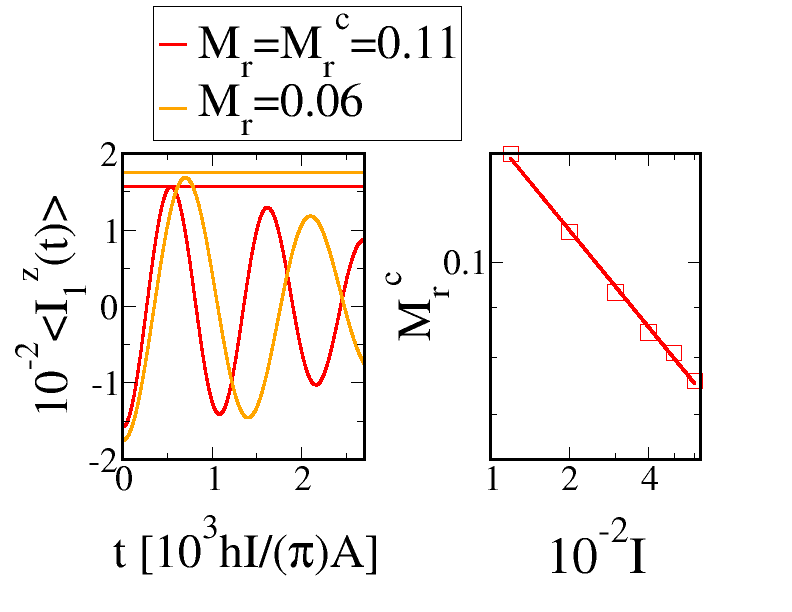}}
\end{flushright}
\caption{\label{Yes} Left panel: $\langle I_1^z(t)\rangle$ for 
$\Jx/(A/2I)=3.5$, $\Delta=0$ and $I=200$. 
The initial state is 
$\ket{\alpha}=(1/\sqrt{13})\left(2\ket{\Uparrow \Downarrow}
+3\ket{\Downarrow \Uparrow} \right)\ket{M-I,I}$ 
where two values of the magnetization $M_r=M/(2I+1)$ are considered; 
$M_r=M_r^c=0.11$ corresponds to the critical value at $I=200$. 
The horizontal lines are guides to the eye indicating the value needed for a 
complete reversal of $\langle I_1^z(t)\rangle$.
Right panel: $M_r^c$ versus spin length $I$ for the same initial state as in 
the left panel: The fit results in a power-law decrease
$M_r^c(I)= e^{-0.5} I^{- 0.33}$. Thus, for large enough spin baths antiparallel 
nuclear spin configurations can, to an excellent degree of accuracy, be 
swapped.}
\end{figure}

Surprisingly, this apparently negative result turns out to be related to the
size of the baths in the following sense: Consider an initial state with
the electron spins being in an arbitrary linear combination of 
$\ket{\Uparrow \Downarrow}$, $\ket{\Downarrow \Uparrow}$ and a nuclear
state $\ket{I_1^z,I_2^z}$ fulfilling, say, $I_1^z<0$ and $I_2^z>0$ with
$|I_1^z|<|I_2^z|$, i.e. the ``magnetization'' $M_r:=M/(2I+1)$
is nonzero, $M_r=(|I_2^z|-|I_1^z|)/(2I+1)\neq 0$.
Here we find that for magnetizations larger 
than a certain ``critical'' value $M_r^c$ (slightly depending on the 
electron spin state) the expectation value $\langle I_1^z (t=0)\rangle$ is, to a
an excellent degree of accuracy, completely
reversed. As a representative example, in the left panel of Fig.~\ref{Yes}
we plot the quantity  $\langle I_1^z(t)\rangle$ for $I=200$ with, as before,
$\ket{\alpha_e}=(1/\sqrt{13})\left(2\ket{\Uparrow \Downarrow}
+3\ket{\Downarrow \Uparrow}\right)$ as initial electron spin state.
We consider two different initial nuclear states 
$\ket{\alpha_n}=\ket{M-I,I}$, where the corresponding value of $M_r$
is in one case exactly at, in the other case lower than the critical 
$M_r^c$ at the given nuclear spin length. In the latter case the reversal
of $\langle I_1^z(t)\rangle$ is slightly incomplete.

\begin{figure}
\begin{flushright}
\resizebox{\linewidth}{!}{
\includegraphics{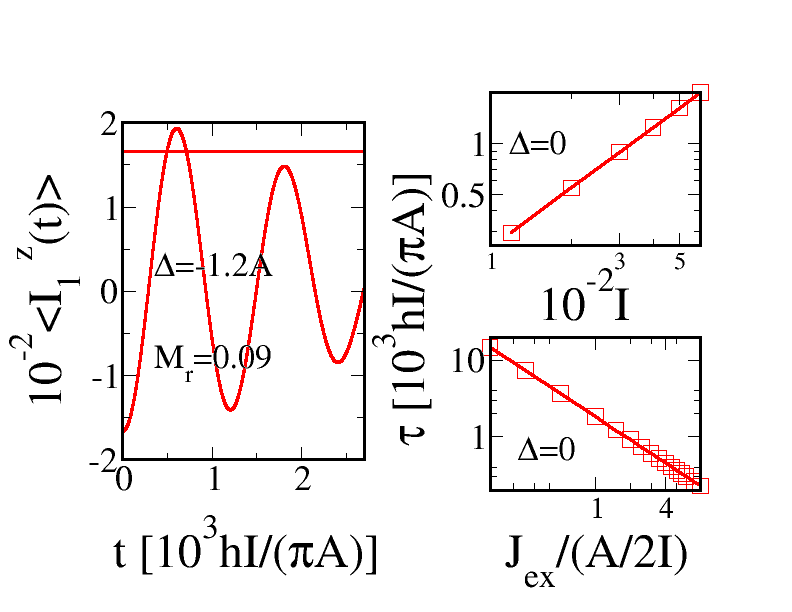}}
\end{flushright}
\caption{\label{Parameters} Left panel: Reversal of $\langle I_1^z(t)\rangle$
for the same situation as in Fig.~\ref{Yes} left panel, but
$\Delta=-1.2A$ and $M_r=0.09$: Breaking inversion symmetry facilitates
swapping the nuclear spin polarizations.
Right panels: Swap time $\tau$ at $\Delta=0$ and $\Jx/(A/2I)=3.5$
as a function of $I$ (upper panel), and at $I=200$ as a function of 
$(\Jx/(A/2I))$ (lower panel). In the upper case we find a power law as
$\tau=e^{0.015}I^{1.19} \left(hI/\pi A\right)$, in the lower case
$\tau=e^{7.51} \left[\Jx/(A/2I)\right]^{-1.003} \left(hI/\pi A\right)$.
}
\end{figure}

It is now a key observation that $M_r^c(I)$ \textit{strongly decreases}
with increasing spin length $I$. This is demonstrated in
the right panel of Fig.~\ref{Yes} for spin lengths
up to $I=600$, where a clear power law scaling is found:
\begin{equation}
M_r^c\approx e^{-\frac{1}{2}} \cdot I^{-\frac{1}{3}}
\end{equation}
Hence, for a large enough value of $I$ the magnetization $M_r^c$ will 
be so close to zero that, up to irrelevant corrections, antiparallel 
nuclear spin configurations can indeed be swapped. To give a quantitative 
example, for $I=10^6$ (as typical
for the nuclear spin bath of GaAs quantum dots) the above power law leads
to $M_r^c=0.006$ implying that baths with 
$I_1^z=-0.988 I_2^z$ can be swapped. Finally, in the bottom panel of 
Fig.~\ref{No} 
we plot  $\langle I_1^z(t)\rangle$ and  $\langle I_2^z(t)\rangle$
for $I=1500$ (which is about the largest system size accessible to our
numerics): Obviously we are very close to a full swap.

Moreover, the performance of such a swap process can significantly be
further improved by departing from the symmetric case $\Delta=A_1-A_2=0$, 
i.e. considering different geometries for the two quantum dots:
The left panel of Fig.~\ref{Parameters} shows $\langle I_1^z(t)\rangle$
for the same situation as in the left panel of Fig.~\ref{Yes}, but
$\Delta=-1.2A$ and $M_r=0.09$ (which is lower than $M_r^c=0.11$ found before
for $\Delta=0$). As seen,  $\langle I_1^z (t=0)\rangle$ is still fully reversed.
Interestingly, this result turns out to be rather independent of the
precise value of  $\Delta\neq 0$ (including its sign), suggesting that
the observed increase of ``swap performance'' goes back to some qualitative
change in the dynamical properties. In fact, as shown in Ref.~\cite{ErbSchl10},
the spectrum of inversion symmetric systems exhibit a macroscopically large
subspace of energetically degenerate multiplets.  Although the initial states 
considered throughout this manuscript lie in energy 
quite far away from those degenerate levels, it is an interesting question
to what extent both observations are related.

Finally, in the right panels of Fig. \ref{Parameters} we analyze the duration
$\tau$ of the swap process as a function of the spin length $I$ as well as the
ratio $\Jx/(A/2I)$ for again $\Delta=0$. 
In both cases we find power law dependencies leading for a realistic system size
of $I=10^6$ to a swap time of $\tau$ of a few ten seconds.  

It is well-known that the nuclear bath is not static. The nuclear spins are interacting
through e.g. dipolar and quadrupolar interactions \cite{Coish09}. These could be possible
limitations to the phenomena described above. In order to circumvent the resulting
problems, one would have to use additional techniques like e.g. refocusing \cite{Hu}.

\textit{Entangling the nuclear baths}.--
In order to measure the entanglement between the long bath spins, 
we utilize the (logarithmic) negativity $L$ defined by \cite{Werner, Plenio}
\begin{equation}
L=\log_2\left(  \| \rho_n^{1} \|_1\right)\,,
\end{equation}
where $\|. \|_1$ denotes the trace norm $\|A\|_1=\Tr(\sqrt{A^+ A})$, and 
$\rho_n^{1}$ is the partial transpose of $\rho_n$ with respect to the 
first spin $\vec{I}_1$. Since
\begin{equation}
\label{Neg}
\| \rho_n^1 \|_1=1+2 \lvert \sum_i E_i^{<} \rvert\,,
\end{equation}
where $E_i^<$ denote the eigenvalues smaller than zero,
 the negativity essentially measures to what extent the partial transpose 
fails to be positive, indicating non-classical correlations \cite{Peres}. 

\begin{figure}
\begin{flushright}
\resizebox{\linewidth}{!}{
\includegraphics{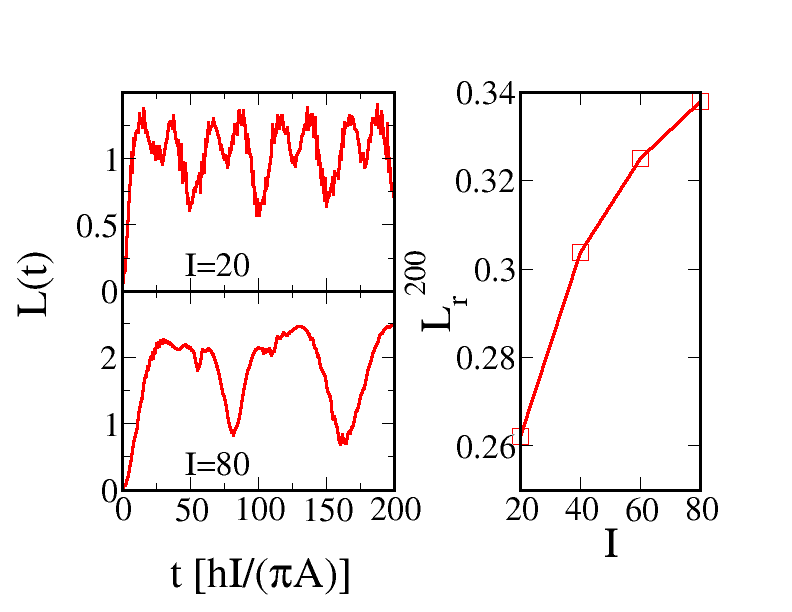}}
\end{flushright}
\caption{\label{PPT_FINAL} Left panels: Negativity $L(t)$
for $I=20, 80$ and $M_r=0.8$, $\Jx/(A/2I)=3.5$, $\Delta=0$. 
Right panel: 
Maximal relative negativity $L_r^{200}$ in the time interval $[0,200](hI/\pi A)$ 
as a function of spin length $I$ for otherwise identical parameters. We 
find a clearly increasing curve, indicating that for large enough sizes, 
the baths can be fully entangled.}
\end{figure}

In order to evaluate the dynamics of the negativity, 
$\rho^1_n$ has to be diagonalized in each time step considered. This is a numerical
effort which restricts us to system sizes somewhat smaller than considered before.
The left panels of Fig.~\ref{PPT_FINAL} show the entanglement dynamics
for two spin lengths $I=20, 80$ at comparatively high polarization
$M_r=0.8$ and $\Delta=0$. The initial state is the same as used before, 
$\ket{\alpha}=(1/\sqrt{13})\left(2\ket{\Uparrow \Downarrow}
+3\ket{\Downarrow \Uparrow} \right) \ket{M-I,I}$.
In both cases the dynamics are rather similar to each other:
A rapid increase of the negativity is followed by a more 
or less regular oscillation around a mean value which increases with the
spin length $I$. In particular, the negativity never returns to zero.

\begin{figure}
\begin{flushright}
\resizebox{\linewidth}{!}{
\includegraphics{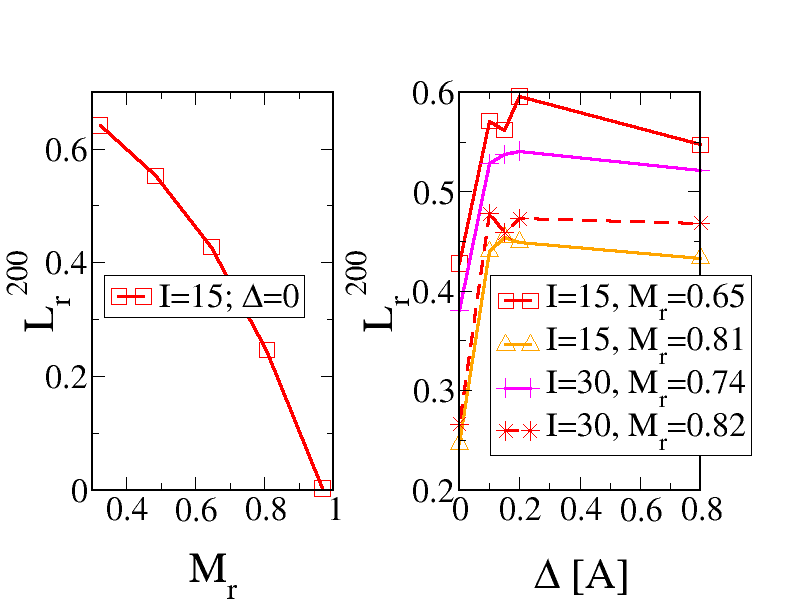}}
\end{flushright}
\caption{\label{MSk} Left panel: Maximal relative negativity 
$L_r^{200}$ versus $M_r$ for $I=15$, $\Delta=0$ and the same initial state as in 
Fig. \ref{PPT_FINAL}. 
Right panel: $L_r^{200}$ as a function of the detuning $\Delta$ for various 
parameters.
The precise value of the detuning is again of no 
particular importance.}
\end{figure}

In order to quantify these observations we introduce a relative
negativity $L_r=L/L_{max}$, where $L_{max}=\log_2 \left( 2I+1 \right)$ is an upper bound of $L$
(cf. Ref.~\cite{Datta}),
and analyze the maximum $L_r^{200}$ of this quantity attained within a fixed
interval $[0,200](hI/\pi A)$.
The results are plotted in the right panel of Fig. \ref{PPT_FINAL}. 
While the spin lengths achievable here are too small to allow
for a quantitatively meaningful fit, the data still shows a significant growth
with increasing $I$ (suggesting, in fact, a power law). This observation implies
that, similarly as for swapping nuclear spin polarizations, also
entangling spin baths benefits from large bath sizes. We note that this effect 
is not due to the
simple growth of the reduced density matrix with increasing $I$ since we
are considering the relative negativity where 
such influences are scaled out. On the other hand, by the same argument,
the maximal relative negativity should decrease with increasing
magnetization at fixed $I$; an example for this behavior is shown
in the left panel of Fig. \ref{MSk}.

In the right panel of Fig. \ref{MSk} we finally demonstrate the influence 
of a non-zero detuning for different spin lengths and magnetizations. 
Similarly to the results regarding a nuclear swap, 
the entanglement is 
enhanced by a non-zero detuning with its precise value being again of
minor importance. This supports the conjecture that the systematic
degeneracy reported in Ref. \cite{ErbSchl10} has a clear dynamical 
signature. Interestingly, breaking the inversion symmetry has stronger influence for higher 
magnetization.


\textit{Conclusions.}--In summary we have studied the spin and entanglement 
dynamics of the nuclear baths in a double quantum dot. Each of the two 
electron spins was considered to interact with an individual bath via 
homogeneous couplings. In order to lower the dimension of the problem, 
both baths have been approximated by long spins. We focused on the 
virtue of the hyperfine interaction and regarded the electron spins 
as an effective coupling between the baths. We demonstrated that it is 
possible to swap them if their size is large enough, and provided strong indication that, 
under the same conditions, it might be even possible to fully entangle 
them. Surprisingly, it turns out to be advantageous to 
use dots of \textit{different} geometry (enabling for $\Delta \neq 0$) to built up the double quantum dot.

\textit{Acknowledgments.}--This work was supported by DFG via SFB 631.

\end{document}